\documentclass[12pt]{article}
\usepackage{graphics}
\input epsf
\usepackage{amssymb}
\parskip=\medskipamount
\def\caE{{\cal E}}

\def\la{\lambda}  
\def\te{\theta}   

\def\k{\kappa}    % kappa
\def\IC{\relax{\rm l\kern-.50 em C}}
\def\IE{\relax{\rm l\kern-.16 em E}}
\def\IK{\relax{\rm l\kern-.18 em K}}
\def\IL{\relax{\rm I\kern-.18 em L}}
\def\IN{\relax{\rm I\kern-.18 em N}}
\def\IR{\relax{\rm I\kern-.18 em R}}
\font\tenfrak=eufm10  \font\sevenfrak=eufm7  \font\fivefrak=eufm5
\newfam\frakfam
\textfont\frakfam=\tenfrak
\scriptfont\frakfam=\sevenfrak
\scriptscriptfont\frakfam=\fivefrak

\newtheorem{proposicion}{Proposition}
\def\wt{\widetilde}
\def\frac#1#2{{#1\over #2}}
\def\fracpd#1#2{\frac{\partial #1}{\partial #2}}
\def\Cos{\mathop{\rm C}\nolimits}    % funcion Coseno
\def\Sin{\mathop{\rm S}\nolimits}    % funcion Seno
\def\r{r}
\def\s{s}
\def\dl{dl}
\def\caP{{\cal P}}
\def\caS{{\cal Q}}
\begin{document}

\title{ The quantum  free particle on spherical and hyperbolic spaces:
 A curvature dependent approach II. }

\author{
Jos\'e F. Cari\~nena$\dagger\,^{a)}$,
Manuel F. Ra\~nada$\dagger\,^{b)}$,
Mariano Santander$\ddagger\,^{c)}$ \\
$\dagger$
   {\sl Departamento de F\'{\i}sica Te\'orica and IUMA, Facultad de Ciencias} \\
   {\sl Universidad de Zaragoza, 50009 Zaragoza, Spain}  \\
$\ddagger$
   {\sl Departamento de F\'{\i}sica Te\'orica, Facultad de Ciencias} \\
   {\sl Universidad de Valladolid,  47011 Valladolid, Spain}
}
\date{Mon, 6 Aug 2012}
%-------------------------------------------------------------------
%%  Version Revisada Revisada Final vRRf
%%  Mon,  6  Aug  2012
%-------------------------------------------------------------------

\maketitle
\begin{abstract}  
This paper is the second part of a study of the quantum
free particle on spherical and hyperbolic spaces by making use of
a curvature-dependent formalism.  Here we study the analogues, 
on the three-dimensional spherical and hyperbolic spaces, $S_\k^3$ 
($\kappa>0$) and $H_\k^3$ ($\kappa<0$), to the standard 
{\itshape spherical waves} in $E^3$. 
The curvature $\k$ is considered as a parameter and for any $\k$ we show how the radial Schr\"odinger equation can be transformed into  a $\k$-dependent Gauss hypergeometric equation that can be considered as a $\k$-deformation of the (spherical) Bessel equation. The specific properties of the spherical waves in the spherical case are studied with great detail. These have a discrete spectrum and their wave functions, which are related with families of orthogonal polynomials (both $\k$-dependent and $\k$-independent), and are explicitly obtained.
\end{abstract}

\begin{quote}
%---------------
{\sl Keywords:}{\enskip}  Quantization. Quantum mechanics on
spaces of constant curvature. Curvature-dependent orthogonal polynomials.

%---------------
{\sl Running title:}{\enskip}
The quantum free particle on three-dimensional spaces with curvature.

%---------------
%%{\it PACS numbers:}
%% {\enskip}03.65.-w, {\enskip}03.65.Ge, {\enskip}02.30.Gp, {\enskip}02.30.Ik
%---------------
%%%  03.65.-w  Quantum mechanics
%%%  03.65.Ge  Solutions of wave equations: bound states
%%%  02.30.Gp  Special functions
%%%  02.30.Ik  Integrable systems

%---------------
{\it MSC Classification:} {\enskip}81Q05, {\enskip}81R12,
{\enskip}81U15, {\enskip}34B24
%---------------
%%% 81Q05 (1991-now) Closed and approximate solutions to the Schr\"odinger, Dirac,
%%% Klein-Gordon and other quantum-mechanical equations
%%% 81R12 (2000-now) Relations with integrable systems
%%% 81U15 (2000-now) Exactly and quasi-solvable systems
%%% 34B24 (1991-now) Sturm-Liouville theory
%---------------
\end{quote}
%%{\vfill}

\footnoterule
{\noindent\small
$^{a)}${\it E-mail address:} {jfc@unizar.es}  \\
$^{b)}${\it E-mail address:} {mfran@unizar.es} \\
$^{c)}${\it E-mail address:} {msn@fta.uva.es}
%---------------
\newpage
%---------------
%%  \tableofcontents

%--------------------------------------
%--------------------------------------
%%  (Section 1.)
\section{Introduction }

This article can be considered as a sequel or continuation of a
previous paper \cite{CaRaSa11a}, which was devoted to the study of
the quantum free particle on two-dimensional spherical and
hyperbolic spaces making use of a formalism that considers the
curvature $\k$ as a parameter. Now, we present a similar analysis
but introducing two changes related with the dimension of the
space and with the states of the quantum free particle we are looking for. 
Now we work in a three-dimensional space, and we look for the states 
analogous to the Euclidean spherical waves, which are determined 
among all free particle states by the condition of being separable in the 
geodesic polar coordinate system.  We follow the approach of \cite{CaRaSa11a}, which
contains the fundamental ideas and motivations, and we also use
the notation, ideas, and results discussed in some related previous
studies \cite{CRSS04NonLin}-\cite{CRS07Jmp}.

There are two articles that are considered of great importance in
the study of mechanical systems  in a spherical geometry (see
\cite{CaRaSa11a} for a more detailed information; we just make here
a quick survey in a rather telegraphic way). Schr\"odinger studied
in 1940 the hydrogen atom in a spherical space \cite{Sch40} and
then other authors studied this problem (hydrogen atom or Kepler
problem) \cite{In41}-\cite{NiSaR99} or other related questions
(as, e.g., the  quantum oscillator on curved spaces)
\cite{BaNe03}-\cite{Gi07}. Higgs studied in  1979 the existence of
dynamical symmetries in a spherical geometry \cite{Hi79} and since
then a certain number of authors have considered
\cite{Le79}-\cite{BaEnHeR09} the problem of the symmetries or some
other properties characterizing the Hamiltonian systems  on curved
spaces (the studies of Schr\"odinger and Higgs were concerned with
a spherical geometry but other authors applied their ideas to the
hyperbolic space). In fact these two problems, the so-called
Bertrand potentials, have been the two problems mainly studied in
curved spaces  (at the two levels, classical and quantum).
Nevertheless in quantum mechanics there are some previous problems
that are of fundamental importance as, for example, the quantum free
particle or the particle in a spherical well.

This article is concerned with the study of the spherical waves for a quantum
free particle on spherical and hyperbolic spaces (an analogous problem
was studied in \cite{CaRaSa11a} but in three dimensions and making use
of $\k$-dependent parallel geodesic coordinates, which affords the 
analogous of plane waves). There are some
questions as, for example, (i)  analysis of some $\k$-dependent
geometric formalisms appropriate for the description of the
dynamics on the spaces with constant curvature $\k$, (ii)
transition from the classical $\k$-dependent system to the quantum
one, (iii) analysis of the Schr\"odinger separability and quantum
superintegrability  on spaces with curvature that have been
discussed in \cite{CaRaSa11a}, and therefore they are now omitted
(or revisited in a very sketchy way). Thus, this paper is mainly
concerned with the exact resolution of the $\k$-dependent
Schr\"odinger equation, existence of bound states, and with 
the associated families of orthogonal polynomials.

In more detail, the plan of the article is as follows:
In Sec. 2 we first study the Lagrangian formalism,  the existence
of Noether symmetries and Noether momenta and then the
$\k$-dependent Hamiltonian and the quantization via Noether
momenta. In Sec. 3, we solve the $\k$-dependent Schr\"odinger
equation and then we analyze with great detail the spherical
$\k>0$ case, writing explicitly the spherical waves on a 3D-sphere 
and discussing their Euclidean limit when the curvature of the sphere goes to 0. 
The study of the hyperbolic $\k<0$ case is only  sketched, but the details displays 
several important differences which would require a separate study. Finally, in Sec. 4 we make some final comments.

%--------------------------------------
%%  (Section 2.)
\section{Geodesic motion,  $\k$-dependent formalism and quantization }

We first present a brief introductory comment on some possible
approaches to the two-dimensional manifolds with constant curvature $\k$: the sphere $S_{\k}^2$
($\k>0$), Euclidean plane $\IE^2$, and hyperbolic plane $H_{\k}^2$
($\k<0$), and then we move to the corresponding three-dimensional
spaces: the sphere $S_{\k}^3$ ($\k>0$), Euclidean space $\IE^3$, and
hyperbolic space $H_{\k}^3$ ($\k<0$).

If we make use of the following $\kappa$-dependent trigonometric
(hyperbolic) functions
$$
 \Cos_{\k}(x) = \cases{
  \cos{\sqrt{\k}\,x}       &if $\k>0$, \cr
  {\quad}  1               &if $\k=0$, \cr
  \cosh\!{\sqrt{-\k}\,x}   &if $\k<0$, \cr}{\qquad}
%---------------
  \Sin_{\k}(x) = \cases{
  \frac{1}{\sqrt{\k}} \sin{\sqrt{\k}\,x}     &if $\k>0$, \cr
  {\quad}   x                                &if $\k=0$, \cr
  \frac{1}{\sqrt{-\k}}\sinh\!{\sqrt{-\k}\,x} &if $\k<0$, \cr}
$$
then the expression of the differential element of distance in
geodesic polar coordinates $(\r,\phi)$ on  $M_{\k}^2 = (S_{\k}^2,
\IE^2, H_{\k}^2)$ can be written as follows
$$
 \dl_{\k}^2 = d\r^2 + \Sin_\k^2(\r)\,d{\phi}^2 \,,
$$
so it reduces to
$$
 \dl_1^2 =    d\r^2 + (\sin^2 \r)\,d{\phi}^2 \,,{\quad}
 \dl_0^2 =    d\r^2 + \r^2\,d{\phi}^2 \,,{\quad}
 \dl_{-1}^2 = d\r^2 + (\sinh^2\r)\,d{\phi}^2\,,
$$
in the three particular cases $\k=1, 0, -1$ of the unit sphere, the Euclidean
plane, and the `unit` Lobachewski plane. 
If we make use of this formalism then the Lagrangian of
the geodesic motion (free particle) on $M_{\k}^2$ is given by
\cite{CRS07Jmp,RaSa02b,RaSa03}
\begin{equation}
 \IL(\k) = (\frac{1}{2})\left(v_\r^2 + \Sin_\k^2(\r) v_\phi^2\right) \,.
\end{equation}
Now if we consider the $\k$-dependent change $\r
\to\,\s=\Sin_\k(\r)$ then the Lagrangian $\IL(\k)$ becomes
$$
 L(\k) = \frac{1}{2}\,\Bigl(\frac{v_{\s}^2}{1 - \k\,\s^2} + \s^2v_\phi^2 \Bigr)\,,
$$
and, if we change to `cartesian coordinates for $s$' defined as $x=\s \cos\phi, y=\s \cos\phi$, we arrive to
$$
 L(\k)  = \frac{1}{2}\,\Bigl(\frac{1}{1 - \k\,\s^2} \Bigr)
 \Bigl[\,v_x^2 + v_y^2 - \k\,(x v_y - y v_x)^2 \,\Bigr]  \,,{\quad}
 \s^2 = x^2+y^2\,,
$$
that is the Lagrangian studied in Ref.
\cite{CaRaSa11a,CRS07AnPh2,CRS07Jmp} (the relation of this
Lagrangian with the Lagrangian of Higgs is also discussed in
\cite{CaRaSa11a,CRS07Jmp}).

We notice that in the sphere case, in addition to the usual geodesic polar coordinate singularity at the origin (the `North pole') $r=0$, which passes to the coordinate $\s$ at $\s=0$, the chart $(\s,\phi)$ covers {\itshape only the upper hemisphere} $0<r<\pi/(2\sqrt{\k})$ as the coordinate $\s$ ceases to be related to $r$ on a one-to-one basis at the equator $r=\pi/(2\sqrt{\k})$, where $\Sin_\k(\r)$ reaches its maximum; however the lower hemisphere $\pi/(2\sqrt{\k})<r<\pi/\sqrt{\k}$ can be also covered by another chart, with $s$ still given by $\s=\Sin_\k(\r)$ with a singularity at $\s=0$, which on the lower hemisphere corresponds to $r=\pi/\sqrt{\k}$, the point antipodal to the origin (the `South pole').

%--------------------------------------
%%  (sub-Section 2.1)
\subsection{Lagrangian formalism,  Noether symmetries and Noether momenta }

Let us start with the following expression for the differential
element of distance $\dl_{\k}$  in the family $M_{\k}^3 =
(S_{\k}^3, \IE^3, H_{\k}^3)$  of three-dimensional spaces with
constant curvature $\k$ written in $(\s,\te,\phi)$ coordinates (recall $\s$ is not the geodesic radial coordinate):
\begin{equation}
 \dl_{\k}^2 = \frac{d\s^2}{1 - \k \, \s^2} + \s^2\,d\te^2
 + \s^2\sin^2 \te\,d\phi^2  \,,
\end{equation}
reducing in the  particular cases of unit sphere,  Euclidean
plane, and  `unit' Lobachewski plane to
%---------------
\begin{eqnarray*}
 \dl_1^2  &=&  \frac{d\s^2}{1 -\, \s^2} + \s^2\,(d\te^2 + \sin^2 \te\,d\phi^2) \,, \cr
 \dl_0^2  &=&  d\s^2  + \s^2\,(d\te^2 + \sin^2 \te\,d\phi^2)    \,, \cr
 \dl_{-1}^2 &=& \frac{d\s^2}{1 +\, \s^2} + \s^2\,(d\te^2 + \sin^2 \te\, \phi^2) \,. 
\end{eqnarray*}

Then the following six vector fields
%---------------
\begin{eqnarray*}
  X_{1}(\k) &=& \sqrt{1 - \k \, \s^2} \,\Bigl[(\sin{\te}\cos{\phi})\,\fracpd{}{\s}
  + \frac{1}{\s} \bigl[(\cos{\te}\cos{\phi})\,\fracpd{}{\te}
  - (\frac{\sin{\phi}}{\sin\te})\,\fracpd{}{\phi}  \bigr] \Bigr]  \,,\cr
  X_{2}(\k) &=& \sqrt{1 - \k \, \s^2} \,\Bigl[ (\sin{\te}\sin{\phi})\,\fracpd{}{\s}
  + \frac{1}{\s} \bigl[(\cos{\te}\sin{\phi})\,\fracpd{}{\te}
  + (\frac{\cos{\phi}}{\sin\te})\,\fracpd{}{\phi} \bigr] \Bigr]\,,\cr
  X_{3}(\k) &=& \sqrt{1 - \k \, \s^2} \,\Bigl[\,(\cos{\te})\,\fracpd{}{\s}
  - \frac{1}{\s} \sin{\te}\,\fracpd{}{\te}   \Bigr]\,,
\end{eqnarray*}
and
$$
  Y_1 =   -\sin\phi\,\fracpd{}{\te}  - (\frac{\cos{\phi}}{\tan \te})\, 
        \fracpd{}{\phi} \,,{\qquad}
  Y_2 =   \cos\phi\,\fracpd{}{\te}  - (\frac{\sin{\phi}}{\tan \te})\,  
        \fracpd{}{\phi} \,,{\qquad}
  Y_3 =   \fracpd{}{\phi} \,,
$$
are  Killing vector fields, that is, the infinitesimal generators
of isometries  of the $\k$-dependent metric $\dl_{\k}^2$.

The Lie brackets of the vector fields $X_i(\k)$ are given by
$$
 [X_2(\k),X_1(\k)] =  \la_\k Y_3 \,,{\quad}
 [X_3(\k),X_2(\k)] =  \la_\k Y_1 \,,{\quad}
 [X_1(\k),X_3(\k)] =  \la_\k Y_2  \,,
$$
with $\la_\k$ given by $\la_\k = \k\sqrt{1-\k\,\s^2}$.
The other Lie brackets are $\k$-independent and similar to the Lie
brackets of the Euclidean case; that is
$$
 [Y_2\,,Y_1] =  Y_3 \,,{\quad}  [Y_3\,,Y_2] =  Y_1 \,,{\quad}
 [Y_1\,,Y_3] =  Y_2 \,,
$$
and so on. All these Killing vector fields close a Lie algebra
that is isomorphic to the Lie algebra of the group of isometries
(either $SO(4), ISO(3), SO(1,3)$) of the spherical, Euclidean or 
hyperbolic spaces depending of the
sign of $\k$. Notice that only when $\k=0$ (Euclidean space),
the vector fields $X_i(\k)$, $i=1,2,3$, will commute between
themselves.

 Now, let us consider the geodesic motion on $M_{\k}^3$, that is, the dynamics
determined by a Lagrangian $L$, which reduces to the $\k$-dependent
kinetic term $T(\k)$ without a potential
\begin{equation}
 L = T(\k) = (\frac{1}{2})\,\Bigl(\frac{v_\s^2}{1 - \k\,\s^2} + \s^2\,v_{\te}^2
 + \s^2\sin^2 \te\,v_{\phi}^2\Bigr)    \,,
\end{equation}
where the parameter $\k$ can take both positive and negative
values. We already mentioned that in the spherical case  the coordinate chart we are dealing with covers only the `upper' 
half-sphere; we see that this Lagrangian becomes singular at the `equator' where $r=\frac{\pi}{2\sqrt{\k}}$, and hence $1 - \k\,\s^2=0$, 
so in this case 
the study of the dynamics will be restricted to the interior of the interval
$0<\s<1/\sqrt{\k}$ which corresponds to the upper half sphere.

The Lagrangian $L =  T(\k)$ is invariant under the action of the
the   $\k$-dependent vector fields $X_i(\k)$ and $Y_i$, $i=1,2,3$, in
the sense that, if we denote by  $X_i^t(\k)$ and $Y_i^t$ the natural
lift to the tangent bundle (phase space $TM_\k^3$ with $M_\k^3$ representing
$S_\k^3$, $\IE^3$, or $H_\k^3$) of the vector fields $X_i(\k)$ and $Y_i$, $ i=1,2,3$, then the Lie derivatives of $T(\k)$ vanish, that is
$$
 X_i^t(\k)\bigl(T(\k)\bigr) = 0 \,,{\quad} Y_i^t\bigl(T(\k)\bigr) = 0 \,,
 {\quad} i=1,2,3.
$$
They represent six exact Noether symmetries for the geodesic motion.
If we denote by $\te_L$ the Lagrangian one-form
\begin{eqnarray*}
 \te_L &=& \Bigl(\fracpd{L}{v_\s}\Bigr)\,d\s + \Bigl(\fracpd{L}{v_\te}\Bigr)\,d\te    + \Bigl(\fracpd{L}{v_\phi}\Bigr)\,d\phi  \cr
  &=& \Bigl(\frac{v_\s}{1 - \k \, \s^2}\Bigr) \,d\s + \s^2 v_{\te}\,d\te
  + \s^2\sin^2\te\,v_{\phi}\,d\phi  \,,
\end{eqnarray*}
then the associated Noether constants of the motion are given by the following:
\begin{itemize}
\item[(P)] The three functions $P_1(\k)$, $P_2(\k)$, and $P_3(\k)$, defined as
$$
 P_i(\k) = i\bigl(X_i^t(\k)\bigr)\te_L \,,\quad   i=1,2,3,
$$
that are $\k$-dependent and  given by
\begin{eqnarray*}
 P_1(\k) &=& (\sin{\te}\cos\phi)\,\frac{v_\s}{\sqrt{1 - \k\,\s^2}} + (\s\sqrt{1 - \k\,\s^2})
   \bigl[(\cos{\te}\cos{\phi})\,v_\te - (\sin{\te}\sin{\phi})\,v_\phi\bigr]\,,\cr
 P_2(\k) &=& (\sin{\te}\sin\phi)\,\frac{v_\s}{\sqrt{1 - \k\,\s^2}} + (\s\sqrt{1- \k\,\s^2})
   \bigl[(\cos{\te}\sin{\phi})\,v_\te + (\sin{\te}\cos{\phi})\,v_\phi\bigr] \,,\cr
 P_3(\k) &=& (\cos{\te})\,\frac{v_\s}{\sqrt{1 - \k\,\s^2}} -  (\s\sqrt{1 - \k\,\s^2})\sin{\te}\,v_\te \,.
\end{eqnarray*}
\item[(J)] The three functions $J_1$, $J_2$, and $J_3$, defined as
$$
 J_i(\k) = i\bigl(Y_i^t\bigr)\te_L \,,\quad   i=1,2,3,
$$
that are $\k$-independent functions and  given by
\begin{eqnarray*}
   J_1  &=&-\,\s^2(\sin\phi\,v_\te + \sin\te\cos\te\cos\phi\,v_\phi) \,,\cr
   J_2  &=&  \s^2 (\cos\phi\,v_\te - \sin\te\cos\te\sin\phi\,v_\phi) \,,\cr
   J_3  &=&  \s^2\sin^2{\te}\,v_\phi\,.
\end{eqnarray*}
\end{itemize}

%--------------------------------------
%%  (sub-Section 2.2)
\subsection{$\k$-dependent Hamiltonian and Quantization  } \label{Sec2.2}

The standard method for the quantization of a Hamiltonian on a Riemannian manifold is to make use of the Laplace-Beltrami operator for the free part (kinetic energy) of the Hamiltonian.   Nevertheless we recall that the standard procedure in a Euclidean space is to first quantize the momenta as self-adjoint operators and then, making use of the quantum momenta, to obtain the quantum version of the Hamiltonian.  Our idea is to translate this momentum-approach to the case of spaces with curvature $\k$ but changing the quantization of the canonical momenta by the quantization of the Noether momenta which are taken as the basic objects  (this is one of the reasons why we have studied the properties of the Killing vectors and Noether momenta with great detail). So, we present the quantization of the system in two steps:  (i) quantization of the Noether momenta as self-adjoint operators and then (ii) quantization of the Hamiltonain making use of the quantum Noether momenta.

The Legendre transformation $(\s,\te,\phi, v_\s,v_\te,v_\phi) \to
(\s,\te,\phi, p_\s,p_\te,p_\phi)$ is given by
$$
  p_\s    = \frac{v_\s}{1 - \k\,\s^2} \,,{\quad}
  p_\te  = \s^2\,v_{\te} \,,{\quad}
  p_\phi = \s^2\sin^2 \te\,v_{\phi} \,,
$$
so that the expression of the $\k$-dependent Hamiltonian turns out to be
\begin{equation}
  H(\k) = \bigl(\frac{1}{2}\bigr)\,\Bigl[(1 - \k\,\s^2)\,p_\s^2
  + \frac{1}{\s^2}\,(p_{\te}^2 + \frac{p_{\phi}^2}{\sin^2\te\,})\Bigr] \,.
\label{Hclas}
\end{equation}
The six Noether momenta become
%---------------
\begin{eqnarray*}
 P_1(\k) &=& \sqrt{1 - \k\,\s^2} \,\Bigl[(\sin{\te}\cos\phi)\,p_\s + \frac{1}{\s}\,
   \bigl[(\cos{\te}\cos{\phi})\,p_\te - (\frac{\sin{\phi}}{\sin{\te}})\,p_\phi\bigr]\Bigr]\,,\cr
 P_2(\k) &=& \sqrt{1 - \k\,\s^2} \,\Bigl[(\sin{\te}\sin\phi)\,p_\s + \frac{1}{\s}\,
   \bigl[(\cos{\te}\sin{\phi})\,p_\te + (\frac{\cos{\phi}}{\sin{\te}})\,p_\phi\bigr] \Bigr]\,,\cr
 P_3(\k) &=& \sqrt{1 - \k\,\s^2} \,\Bigl[(\cos{\te})\,p_\s - \frac{1}{\s}\,\sin{\te}\,p_\te \Bigr]\,,
\end{eqnarray*}
and
$$
   J_1 = -\,\sin\phi\,p_\te - (\frac{\cos{\phi}}{\tan{\te}})\,p_\phi\,,\quad
   J_2 =   \cos\phi\,p_\te - (\frac{\sin{\phi}}{\tan{\te}})\,p_\phi \,,\quad
   J_3 =  p_\phi        \,,
$$
with Poisson brackets
$$
 \{P_i(\k)\,,H(\k)\} = 0  \,,{\quad} \{J_i\,,H(\k)\} = 0  \,,{\quad} i=1,2,3,
$$
and
$$
 \{P_1(\k)\,,P_2(\k)\} =  \k\,J_3 \,,{\quad}
 \{P_2(\k)\,,P_3(\k)\} =  \k\,J_1 \,,{\quad}
 \{P_3(\k)\,,P_1(\k)\} =  \k\,J_2 \,.
$$
The other Poisson brackets are similar to the Poisson brackets of
the Euclidean case; that is,
$$
 \{P_1(\k)\,,J_1\} =  0 \,,{\quad}
 \{P_1(\k)\,,J_2\} =  P_3(\k) \,,{\quad}
 \{P_1(\k)\,,J_3\} =  -P_2(\k) \,,
$$
and so on. Note the change of the order in the Poisson brackets.
This is motivated because of the property $[X_f,X_g] =
-\,X_{\{f,g\}}$; that is, the map $[\,,\,]\to\{\,,\,\}$ is a Lie
algebra isomorphism but with a change of the sign.

Making use of this formalism, the Hamiltonian of the $\k$-dependent oscillator can be rewritten as follows
\begin{equation}
  H(\k) = \bigl(\frac{1}{2 m}\bigr) \Bigl[ P_1^2 + P_2^2 +P_3^2 
  + \k\,(J_1^2 +J_2^2+J_3^2) \Bigr]
\end{equation}

%--------------------
%%  {Proposicion 1}
\begin{proposicion} \label {Proposicion 1}
The only  measure that is invariant under the action of
the three vector fields $X_i(\k)$ and the three vector fields
$Y_i$,   is given in coordinates $(\s, \te, \phi)$ and up to a constant factor by 
$$
  d\mu_\k =  \Bigl(\frac{\s^2\,\sin\te}{\sqrt{1-\k\,\s^2}}\Bigr)\,d\s\,d\te\,d\phi \,.  
$$
\end{proposicion}
%--------------------
This property is proved as follows. The most general expression
for a volume three-form is given by
$$
  \omega = \mu(\s,\te,\phi)\,d\s{\wedge}d\te{\wedge}d\phi
$$
where $\mu(\s,\te,\phi)$ is  a differentiable function to be determined.
Then the conditions
$$
 {\cal L}_{Y_i} \,d  \omega = 0 \,,{\quad}
 {\cal L}_{Y_i} \,d  \omega = 0 \,,{\quad} i=1,2,3,
$$
lead to the following value for the function $\mu$:
$$
  \mu = K\,\Bigl(\frac{\s^2\,\sin\te}{\sqrt{1-\k\,\s^2}}\Bigr)  \,,
$$
where $K$ is an arbitrary constant. Assuming $K=1$ we obtain $d\mu_\k$.

  This  property suggests the appropriate procedure for
obtaining the quantization of the Hamiltonian $H(\k)$.
The idea is to work with functions and linear operators defined on
the space obtained by considering the three-dimensional space
endowed with the measure $d\mu_\k$.  This means, in the
first place, that the operators $\widehat{P_1}$,  $\widehat{P_2}$,
and $\widehat{P_3}$, representing the quantum version of of the
Noether  momenta $P_1$, $P_2$, an $P_3$ must be self-adjoint
not in the standard space $L^3(\IR^3)$ but in the space
$L^2(\IR^3,d\mu_\k)$.
If we assume the following correspondence:
\begin{eqnarray*}
 P_1 \ \mapsto\  \widehat{P_1}&=&  -\,i\,\hbar\,
 \sqrt{1 - \k \, \s^2} \,\Bigl[ (\sin{\te}\cos{\phi})\,\fracpd{}{\s}
  + \frac{1}{\s} \bigl[(\cos{\te}\cos{\phi})\,\fracpd{}{\te}
  - (\frac{\sin{\phi}}{\sin\te})\,\fracpd{}{\phi}  \bigr] \Bigr] \,,\cr
 P_2 \ \mapsto\  \widehat{P_2}&=& -\,i\,\hbar\,
\sqrt{1 - \k \, \s^2} \,\Bigl[ (\sin{\te}\sin{\phi})\,\fracpd{}{\s}
  + \frac{1}{\s} \bigl[(\cos{\te}\sin{\phi})\,\fracpd{}{\te}
  + (\frac{\cos{\phi}}{\sin\te})\,\fracpd{}{\phi} \bigr] \Bigr]   \,,\cr
 P_3 \ \mapsto\  \widehat{P_3}&=& -\,i\,\hbar\,
 \sqrt{1 - \k \, \s^2} \,\Bigl[\,(\cos{\te})\,\fracpd{}{\s}
  - \frac{1}{\s} \sin{\te}\,\fracpd{}{\te} \Bigr]  \,,
\end{eqnarray*}
and
\begin{eqnarray*}
 J_1 \ \mapsto\  \widehat{J_1}&=&   \,i\,\hbar\,\Bigl[\,\sin\phi\,\fracpd{}{\te}
      + (\frac{\cos{\phi}}{\tan \te})\,\fracpd{}{\phi}\,\Bigr] \,,\cr
 J_2 \ \mapsto\  \widehat{J_2}&=&  -\,i\,\hbar\,\Bigl[\,\cos\phi\,\fracpd{}{\te}
      - (\frac{\sin{\phi}}{\tan \te})\,\fracpd{}{\phi}\,\Bigr]   \,,\cr
 J_3 \ \mapsto\  \widehat{J_3}&=&   -\,i\,\hbar\, \fracpd{}{\phi} \,,
\end{eqnarray*}
then we have
$$
 P_1^2 + P_2^2 +P_3^2  \mapsto  -\,\hbar^2\biggl[ (1 - \k\,\s^2)\Bigl[\,\fracpd{^2}{\s^2}
 + \frac{1}{\s^2}\,\Bigl(\,\fracpd{^2}{\te^2}+ \frac{1}{\sin^2\te}\,\fracpd{^2}{\phi^2}
 + \frac{1}{\tan\te}\fracpd{}{\te}\Bigr)\Bigr]+ \frac{2 - 3\k\,\s^2}{r}\,\fracpd{}{\s}   \biggr] \,,
$$
and
$$
 J_1^2 +J_2^2+J_3^2  \mapsto  -\,\hbar^2 \Bigl[\, \fracpd{^2}{\te^2}  +\frac{1}{\sin^2\te}\fracpd{^2}{\phi^2}+\frac{1}{\tan\te}\fracpd{}{\te}\,\Bigr]\,,
$$
in such a way that the quantum  Hamiltonian   $\widehat{H}(\k)$
\begin{equation}
 \widehat{H}(\k) =\bigl(\frac{1}{2 m}\bigr)
 \Bigl[ \widehat{P_1}^2 + \widehat{P_2}^2 +\widehat{P_3}^2
 + \k\,(\widehat{J_1}^2 +\widehat{J_2}^2+\widehat{J_3}^2)\Bigr] \,,
\label{Hq1}
\end{equation}
is represented by the following differential operator:
\begin{equation}
 \widehat{H}(\k) = - \frac{\hbar^2}{2 m}\, \Bigl[(1 - \k\,\s^2)\,\fracpd{^2}{\s^2}
 + \frac{2 - 3\k\,\s^2}{\s}\,\fracpd{}{\s} + \frac{1}{\s^2}\,\Bigl(
 \fracpd{^2}{\te^2} + \frac{1}{\sin^2\te}\fracpd{^2}{\phi^2} + \frac{1}{\tan\te}\fracpd{}{\te}\Bigr) \Bigr]    \,.
\label{Hq2}
\end{equation}
We note that this operator is self-adjoint with respect the measure $d\mu_\k$ and also that it satisfies the appropriate Euclidean limit (in this limit $s$ goes to the Euclidean radial coordinate $r$, so to conform with the standard Euclidean usage we write $r$ in this expression):
$$
 \lim_{\k\to 0} \widehat{H}(\k) =   - \frac{\hbar^2}{2 m}\,
 \Bigl[\,\fracpd{^2}{r^2} + \frac{2}{r}\,\fracpd{}{r} + \frac{1}{r^2}\,
 \Bigl( \fracpd{^2}{\te^2}+\frac{1}{\sin^2\te}\fracpd{^2}{\phi^2}
 +\frac{1}{\tan\te}\fracpd{}{\te}\Bigr) \Bigr]    \,.
$$

Note that if we write $\widehat{H}(\k)= \widehat{H}_1  + \widehat{H}_2 + \widehat{H}_3$ with  $\widehat{H}_i = \widehat{P_i}^2 + \k\widehat{J_i}^2$, $i=1,2,3$,  then $ \bigl[ \widehat{H}_i\,, \widehat{H}_j \bigr] \ne  0$, $i\ne j$, because of the $\k$-dependent terms. Nevertheless $ \bigl[ \widehat{H}_i\,, \widehat{H}_{jk} \bigr] =  0$, $\widehat{H}_{jk}=\widehat{H}_j  + \widehat{H}_k$, $i\ne j\ne k$, so that   $\widehat{H}(\k)$ can be written as sum of two operators that conmute in several different ways.
Finally we also note that $\widehat{H}(\k)$ can also be written as
$$
 \widehat{H}(\k) = \widehat{H}_P +   \k\, \widehat{H}_J  \,,{\quad}
 \bigl[ \widehat{H}_P\,, \widehat{H}_J\bigr] =  0\,,{\quad}
  \widehat{H}_P = \widehat{P}^2 \,,{\quad}
  \widehat{H}_J = \widehat{J}^2 \,,
$$
that corresponds to the approach considered in this paper.

We close this section with the following observations: 

\begin{enumerate}  

\item 
Only for reference, we mention that had we used the polar geodesic coordinates $(r, \te, \phi)$, then the Hamiltonian would have be represented by the following differential operator:
\begin{equation}
 \widehat{H}(\k) =   - \frac{\hbar^2}{2 m}\, \Bigl[\, \frac{1}{S_{\kappa}^2(r)} 
 \frac{d}{d r}\left(S_{\kappa}^2(r) \frac{d}{d r}\right) + \frac{1}{S_{\kappa}^2(r)}\,  
 \Bigl( \fracpd{^2}{\te^2}+\frac{1}{\sin^2\te}\fracpd{^2}{\phi^2}
 +  \frac{1}{\tan\te}\fracpd{}{\te}\Bigr) \Bigr]    \,. 
\label{HamGeodCoord}
\end{equation}

\item  The measure $d\mu_\k$ was  introduced as the unique measure (up to a multiplicative constant) invariant under the Killing vectors. We have verified that it  coincides with the corresponding  Riemann volume in a space with curvature $\k$. 

\item  The Noether momenta quantization procedure was motivated in the first paragraph of section \ref{Sec2.2}. At this point we must clearly state that the final result (that is, the expression of the Hamiltonian operator) coincides with the one that would be obtained making use of the  Laplace-Beltrami  quantization. In fact, the fundamental point for the validity of our approach was that the Hamiltonian is the quadratic Casimir of the isometry algebra; this leads to the Laplace-Beltrami operator. 

\end{enumerate}

%--------------------------------------
%% (Section 3.)
\section{$\k$-dependent Schr\"odinger equation }

The Schr\"odinger equation:
\begin{equation}
 \widehat{H}(\k)\,\Psi = E\,\Psi\,,  {\quad} E = E_{P}+ \k\,E_J \,,
\label{EcSchHk}
\end{equation}
leads in the coordinates $(\s, \te, \phi)$ we are using to the following $\k$-dependent differential equation:
\begin{equation}
 - \frac{\hbar^2}{2 m}\, \Bigl[(1 - \k\,\s^2)\,\fracpd{^2}{\s^2}
 + \frac{2 - 3\k\,\s^2}{\s}\,\fracpd{}{\s} + \frac{1}{\s^2}\,\Bigl(
 \fracpd{^2}{\te^2}+\frac{1}{\sin^2\te}\fracpd{^2}{\phi^2}+\frac{1}{\tan\te}\fracpd{}{\te}\Bigr) \Bigr]     \,\Psi =   E\,\Psi  \,.
\end{equation}
Thus, if we assume that $\Psi(\s,\te,\phi)$ can be factorized in the form
\begin{equation}
 \Psi(\s,\te,\phi) = R(\s) \,Y_{Lm}(\te,\phi) \,,
\label{sFactorization}
\end{equation}
where $R$ is a function of $\s$ and $Y_{Lm}(\te,\phi)$ are the
standard $\k$-independent spherical harmonics
 $$
 \Bigl( \fracpd{^2}{\te^2} +\frac{1}{\sin^2\te}\fracpd{^2}{\phi^2}
 +\frac{1}{\tan\te}\fracpd{}{\te} \Bigr)\,Y_{Lm} = -\,L(L+1)\,Y_{Lm}
$$
then we arrive to the following $\k$-dependent radial equation:
$$
 - \frac{\hbar^2}{2 m}\, \Bigl[(1 - \k\,\s^2)\,\frac{d^2}{d\s^2}
 + \frac{2 - 3\k\,\s^2}{\s}\,\frac{d}{d\s} - \frac{L(L+1)}{\s^2}\Bigr]\,R= E\,R
  \,,{\quad}   R = R(\s) \,.
$$
that can be rewritten in the form
\begin{equation}
 \Bigl[(1 - \k\,\s^2)\,\frac{d^2}{d\s^2}  + \frac{2 - 
 3\k\,\s^2}{\s}\,\frac{d}{d\s} - \frac{L(L+1)}{\s^2} \, + \caE^2\Bigr]\,R=0 \,,
\label{EcR(s)}
\end{equation}
where $\caE$ is defined for any value of the curvature so that it bears with the energy the same relation as the modulus of the wave vector $k$ has with $E$ in the Euclidean case: 
$$
  \caE^2 = \frac{2 m E}{\hbar^2}  \,.
$$
If we adimensionalyze the radial variable $\s$ and the curvature $\k$ to the
new adimensional variables
$$
  \rho =\caE\, \s \,,\quad \wt{\k} =\k/\caE^2  \,,{\quad}
  \k\,\s^2 = \wt{\k}\,\rho^2 \,,
$$
then we arrive to the following equation for the function $R(\rho)$:
\begin{equation}
 {\rho^2} (1 - \wt{\k}\,\rho^2)\,R'' + \rho(2 - 3\wt{\k}\,\rho^2)\,R'
 +\bigl( \rho^2 - L(L+1) \bigr)\,R= 0 \,,
\label{EcR(rho1)}
\end{equation}
that represents a $\k$-dependent deformation of the spherical Bessel differential equation
\begin{equation}
 \rho^2\,R'' + 2{\rho} R' +\bigl[\rho^2 - L(L+1)  \bigr] \,R  = 0 \,.
\label{EcR(rho2)}
\end{equation}
This `deformed' spherical Bessel equation can be solved in power series using the method of Frobenius.
First the function $R$ must be written as follows
$$
  R = \rho^\mu\,f(\rho,\wt{\k}) \,,
$$
and then it is proved that $\mu$ must take one of the two values
$\mu_1 = L$ or $\mu_2=-L-1$.  Choosing $\mu=L$, in order to translate to the unknown function 
$f(\rho)$ the condition $R$ has to satisfy to be well defined at the origin in a simpler form, we arrive at
\begin{equation}
 \rho\,(1 - \wt{\k}\,\rho^2)\,f'' + \bigl[2 (L+1) -  \wt{\k} (2L + 3)\rho^2 ]\,f'
 + \bigl[ 1 -  \wt{\k} L (L+2)\bigr]\,\rho\,f = 0  \,. 
\label{Ecf(rho1)}
\end{equation}

%--------------------------------------
%% (Section 4.)
\section{Spherical $\k>0$ case}

Let us now consider the spherical case $\k>0$.  Before starting, we recall that the coordinates $(\s, \te,\phi)$ we are using cover only the `upper' half the the sphere (where $r$ ranges in the interval $[0, \frac{\pi}{2\sqrt{\k}}]$), so the range of $\s$ is $[0, \frac{1}{\sqrt{\k}}]$, and the range of the new variable $\rho$ is $[0, \frac{1}{\sqrt{\wt\k}}]$. A quite similar coordinate chart (with the same relation with $r$) covers the other half, so at the end our results will cover the whole sphere. 

We will prove that the equation (\ref{Ecf(rho1)}) admits two different types of solutions.

%--------------------------------------
%%  (sub-Section 4.1)
\subsection{Solutions of type I}

Assuming a $\k$-dependent power series for $f$
$$
 f  = \sum_{n=0}^{\infty}\,  f_n \rho^n
 = f_0 + f_1 \rho + f_2 \rho^2 + f_3 \rho^3 + \dots
$$
then the $\k$-dependent recursion relation leads to the vanishing
of all the odd coefficients, $ f_1 = f_3 = f_5 = \dots = 0$,
so that $f$ is a series with only even powers of $\rho$ and a
radius of convergence $R_c$ given by $R_c =
1/\sqrt{\,|\,\wt{\k}\,|\,}$ (determined by the presence of the
second singularity). The even powers dependence suggests to
introduce the new variable $z=\rho^2$ so that the equation becomes
\begin{equation}
  4 z\,(1 - \wt{\k}\,z)\, f_{zz}'' + 2 [ (2L+3)  - 2 \wt{\k} (L+2)z]\,f_z' + [1 - \wt{\k} L (L+2)] f = 0  \,.
\label{Ecf(zk1)}
\end{equation}
 As we have $\wt{\k} > 0$, it is  convenient to complement the previously suggested variable change $z=\rho^2$ with a further last change $t=\wt{\k}\, z$, with the range $[0,1]$ for $t$.
Then the equation (\ref{Ecf(zk1)}) reduces to
\begin{equation}
 t\,(1 - t)\,f_{tt}'' + \Bigl[(L+\frac{3}{2}) - (L+2)\,t\Bigr]\,f_t'
 + \frac{1}{4\wt{\k}}\bigl[1 - \wt{\k}\,L\,(L+2)\, \bigr]\,f  = 0   \,,
\label{Ecf(t)}
\end{equation}
that is, a Gauss hypergeometric equation
$$
 t\,(1 - t)\,f_{tt}'' + [c-(1+a_\k+b_\k)t]\,f_t' - a_\k b_\k f = 0 \,,
$$
with
$$
 c = L+\frac{3}{2}   \,,\qquad
 a_\k + b_\k = L + 1  \,,\qquad
 a_\k b_\k = - \frac{1}{4\wt{\k}}\bigl[1 - \wt{\k}\,L\,(L+2)\, \bigr]
 \,,
$$
and the solution regular at $t=0$ is the hypergeometric function
$$
 f(t,\k) = {}_2 F_1(a_\k,b_\k;c\,; t) \,,\qquad
 {}_2 F_1(a_\k,b_\k;c\,; t) = 1 + \sum_{n=1}^{\infty}\,
 \frac{(a_\k)_n\,(b_\k)_n}{(c)_n} \,\frac{t^n}{n\,!} \,,
$$
with $a_\k$ and $b_\k$ given by
$$
 a_\k = \frac{1}{2}\Bigl[(L+1) \pm B_\k  \Bigr]  \,,{\quad}
 b_\k = \frac{1}{2}\Bigl[(L+1) \mp B_\k  \Bigr] \,,{\quad}
 B_\k = \frac{\sqrt{\wt{\k}(\wt{\k}+1)}}{\wt{\k}}  \,.
$$
The equation has a singularity at $t=1$ that corresponds to
$z=1/\wt{\k}$ (this is, to $\rho=1/\sqrt{\wt{\k}}$ or to $r=\pi/(2\sqrt{{\k}})$).  
If the origin $r=0$ is placed in the `north pole' of the sphere then this singularity is
just at the equator, which is the boundary of the domain covered by the coordinate chart $\s$. 
The property of regularity of the solutions leads to analyze the existence of particular solutions
well defined at this point. 
The polynomial solutions  appear when one of the two $\k$-dependent coefficients, 
$a_\k$ or $b_\k$, coincide with zero or with a negative integer number
$$
 a_\k = -\,n_r\,,\quad {\rm or}\quad
 b_\k = -\,n_r\,,\qquad  n_r=0,1,2,\dots
$$
Then, in this case, we have
$$
  \sqrt{\wt{\k}(\wt{\k}+1)} =  -\wt{\k} \,(2 n_r + L + 1)  \,,
$$
that can be writen as
$$
   \wt{\k} =\k/\caE^2  =  1/((2 n_r + L) ( 2 n_r + L + 2))   \,.
$$
Therefore the coefficient $\caE$ that represents the sphere analogue of the modulus of the wave vector of the spherical wave is
restricted to one of the  values ${\caE}_{{n_r}L}$ given in the discrete set
\begin{equation}
 {\caE}_{{n_r}L}^2 = \k\, (2 n_r + L) ( 2 n_r + L + 2) 
  = \k\, [(2 n_r+L+1)^2-1] \,,
\label{E(nrL)}
\end{equation}
and then the hypergeometric series ${}_2 F_1(a_\k,b_\k,c\,; \wt{\k} z)$ reduces
to a polynomial of degree $n_r$ in the variable $z$.

Coming back to the equation (\ref{EcR(s)}), that was written making use of the 
radial variable $\s$, then the recurrence relation leads to the following values for the even coefficients
$$
 f_2 = \frac{\k L (L+2) - \caE^2}{2 (2 L + 3)}\,\,f_0  \,,{\quad}
 f_4 = \frac{\k (L+2)(L+4) - \caE^2 }{4 (2 L + 5)}\,\,f_2  \,,{\quad}
 f_6 = \frac{\k (L+4)(L+6) - \caE^2 }{6 (2 L + 7)}\,\,f_4 \,,{\quad}  \dots
$$
The result is that $f$, when written as a function of $s$,  is given by the following expression: 
\begin{eqnarray*}
 f &=&  1 + \frac{\k L (L+2) - \caE^2}{2 (2 L + 3)}\,\s^2
  + \frac{[\k L (L+2) - \caE^2][\k (L+2)(L+4) - \caE^2]}{8 (2 L + 3)(2 L + 5)}\,\s^4 \\[6pt]
 &+&      \frac{[\k L (L+2) - \caE^2][\k (L+2)(L+4) - \caE^2][\k (L+4)(L+6) - \caE^2]}{48 (2 L + 3)(2 L + 5)(2 L + 7)}\,\s^6  + \dots
 \end{eqnarray*}
which turns out to be a polynomial of degree $2 n_r$, which will be denoted  as ${\caP}_{{n_r}L}^{\,f}$.

\begin{enumerate}
\item  Suppose that $n_r=0$; then $\caE^2 = \caE_{0L}^2 = \k L (L+2)$
and $f_2=0$. In this particular case the function $f$ reduces to
$$
  {\caP}_{0L}^{\,f}(\s) = 1 \,.
$$
\item  Suppose that $n_r=1$; then $\caE^2 = \caE_{1L}^2 = \k
(L+2)(L+4)$ and $f_4=0$. In this particular case, the function $f$
reduces to the quadratic polynomial 
$$
  {\caP}_{1L}^{\,f}(\s) =  1 - \frac{2 (L+2)}{(2 L + 3)}\, \k\s^2 \,.
$$

\item  Suppose that $n_r=2$; then $\caE^2 = \caE_{2L}^2 = \k
(L+4)(L+6)$ and $f_6=0$.   In this particular case, the function $f$
reduces to the polynomial$$
  {\caP}_{2L}^{\,f}(\s) = 1 - \frac{4 (L+3)}{(2 L + 3)}\, \k\s^2 + \frac{4 (L+3)(L+4)}{(2 L + 3)(2 L + 5)}\, \k^2\s^4\,.
$$

\item  Suppose that $n_r=3$; then $\caE^2 = \caE_{3L}^2 = \k
(L+6)(L+8)$ and $f_8=0$.  In this particular case, the function $f$
reduces to the polynomial
$$
  {\caP}_{3L}^{\,f}(\s)  =   1 -  \frac{6 (L+4)}{(2 L + 3)}\, \k\s^2 + 
  \frac{12 (L+4)(L+5)}{(2 L + 3)(2 L + 5)}\, \k^2\s^4
  - \frac{8 (L+4)(L+5)(L+6)}{(2 L + 3)(2 L + 5)(2 L + 7)}\, \k^3\s^6\,.
$$
\item In the general case for any $n_r$ then $\caE^2 = \caE_{n_rL}^2 = \k\, (2 n_r + L) (2 n_r + L + 2)$, $n_r=0,1,2,\dots$, and  then $f_{2n_r} \ne 0$, but $f_{2n_r+2}=0$. The function $f$ reduces to  a polynomial ${\caP}_{n_rL}^{\,f}$ of degree $2n_r$ in the variable $\s$, with only even powers.

\end{enumerate}

Of course, one should recall that the radial solution $R$,  is the product of $\s^L$ times the polynomial ${\caP}_{n_rL}^{\,f}(\s)$. 
We note that in every polynomial,  the coefficient $c_{2i}$ of
$\s^{2i}$ is proportional to $\k^i$ so that the `direct' Euclidean limit $\k\to 0$ of
all these polynomials is just the unity, that is, ${\caP}_{n_rL}\to 1$
when $\k\to 0$. However, if we come back to the equation (\ref{EcR(s)}) which holds for any value of $\k$, then we find that for $\k=0$ the recurrence relations are precisely
$$
 f_2 = \frac{ - \caE^2}{2 (2 L + 3)}\,\,f_0  \,,{\quad}
 f_4 = \frac{ - \caE^2}{4 (2 L + 5)}\,\,f_2  \,,{\quad}
 f_6 = \frac{ - \caE^2}{6 (2 L + 7)}\,\,f_4  \,,{\quad}\dots
$$
showing that in this case the differential equation does not admit polynomial solutions, and that the relevant solutions should be given by a series involving arbitrarily high powers of $\s$ (starting at $\s^L$). Of course, this is to be expected as in this Euclidean case we already knew that the pertinent solutions of the full radial equations are the spherical Bessel functions, $j_L(k r)$, which after extracting a factor $r^L$ are not a polynomial in the variable $r$. In the next subsection we come back on this question. 

%--------------------
%%  {Proposicion 2}
\begin{proposicion} \label{Prop2}
The eigenfunctions corresponding to distinct eigenvalues of the
Sturm-Liouville problem given by the equation
$$
 a_0  f'' + a_1 f' + a_2 f = 0  \,,
$$
with
$$
 a_0 = \s\,(1 - k\,\s^2) \,,{\quad}  a_1 =  2 (L+1) -  \k (2L + 3) \s^2  \,,{\quad}
 a_2 = \bigl[ \caE^2 -  \k L (L+2)\bigr]\,\s  \,,
$$
together with the appropriate boundary conditions in the points
$\s_1=0$ and $\s_2=1/\sqrt{\k}$  are orthogonal in the interval
$[0,1/\sqrt{\k}]$ with respect to the weight function $q =
\s^{2(L+1)}/\sqrt{1 - \k\,\s^2}$.
\end{proposicion}

{\it Proof:}  This $\k$-dependent
differential equation is not self-adjoint since $a'_0 \ne a_1$, but
it can be reduced to self-adjoint form by making use of an appropriate 
integrating factor so that the equation becomes
\begin{equation}
  \frac{d}{d\s}\Bigl[\,p(\s,\k)\,\frac{df}{d\s}\,\Bigr] +  \la\, q(\s,\k)\,f = 0 \,,
\label{EcSLf(s)} 
\end{equation}
with
$$
  p(\s,\k)=  \s^{2(L+1)} \sqrt{1 - \k\,\s^2}   \,,{\quad}
  q(\s,\k) = \frac{ \s^{2(L+1)}}{ \sqrt{1 - \k\,\s^2}}  \,,{\quad}
  \la = \caE^2 -\k L(L+2)\,.
$$
This is a Sturm-Liouville problem that  is  singular  since the function $p(\s,\k)$
vanish in the two end points $\s_1$ and $\s_2$;  therefore, the appropriate boundary 
conditions are the boundedness of the solutions (and their derivatives) 
at the singular end points. The properties of the Sturm-Liouville problems state that even in
this singular case the eigenfunctions of the problem are orthogonal with
respect to the function $q(\s,\k)$.

The eigenfunctions are just the polynomial solutions ${\caP}_{n_rL}^{\,f}$, $n_r=0,1,2,\dots$ previously obtained that satisfy the
orthogonality relations
$$
  \int_{0}^{1/\sqrt{\k}} {\caP}_{n_{r_1}L}^{\,f}(\s,\k)\,{\caP}_{n_{r_2}L}^{\,f}(\s,\k)\, q(s,\k)\, d\s = 0   \,,\quad n_{r_1}\,\ne\,n_{r_2} \,,\quad \k>0\,,
$$
that can be rewritten as follows
$$
  \int_{0}^{1/\sqrt{\k}} \Bigl[\s^L {\caP}_{n_1L}^{\,f}(\s,\k)\Bigr]\,
  \Bigl[\s^L{\caP}_{n_2L}^{\,f}(\s,\k)\Bigr]
  \Bigl(\frac{\s^2}{\sqrt{1-\k\,\s^2}}\Bigr)\,d\s  = 0   \,,\quad
  n_{r_1}\,\ne\,n_{r_2} \,,\quad \k>0\,.
$$

It is important to note that this last orthogonality relation 
is with respect a $\k$-dependent weight function that coincides
with measure $d\mu_k$ obtained  in  Proposition
 1 for carrying on the quantization.

%--------------------------------------
%%  (sub-Section 4.2)
\subsection{Solutions of type II}

Let us suppose for the function $f(\rho)$ the following factorization:
$$
 f(\rho) = \sqrt{1 - \wt{\k}\,\rho^2}\,\,g(\rho)  \,. 
$$
Then Eq. (\ref{Ecf(rho1)})  becomes  
\begin{equation}
  \rho\,(1 - \wt{\k}\,\rho^2)\,g'' + \bigl[2 (L+1) -  \wt{\k} (2L + 5)\rho^2 ]\,g'
 + \bigl[ 1 -  3\,\wt{\k} - \wt{\k}\,L (L+4)\bigr]\,\rho\,g = 0  \,.
\label{Ecg(rho)}
\end{equation}
Assuming a power series for $g$, 
$$
 g  = \sum_{n=0}^{\infty}\,  g_n \rho^n
 = g_0 + g_1 \rho + g_2 \rho^2 + g_3 \rho^3 + \dots
$$
then the $\k$-dependent recursion relation leads to the vanishing
of all the odd coefficients, $g_1 = g_3 = g_5 =   \dots = 0$,
so that $g$ is a series with only even powers of $\rho$ and a
radius of convergence $R_c$ given by $R_c =
1/\sqrt{\,|\,\wt{\k}\,|\,}$ (determined by the presence of the
second singularity). 
The even powers dependence suggests to
introduce the new variable $z=\rho^2$ so that the equation becomes
\begin{equation}
  4 z\,(1 - \wt{\k}\,z)\, g_{zz}'' + 2 [ (2L+3)  -  2\, \wt{\k} (L+3)\,z]\,g_z' +  \bigl[ 1 -   \wt{\k}\,(3 +\,L (L+4))\bigr]\, g = 0  \,. 
\label{Ecg(z)}
\end{equation}
The change $t= \wt{\k}\,z$ leads to 
\begin{equation}
 t\,(1 - t)\,g_{tt}'' + \Bigl[(L+\frac{3}{2}) - (L+3)\,t\Bigr]\,g_t'
 + \frac{1}{4\wt{\k}}\,\bigl[ 1 -   \wt{\k}\,(3 +\,L (L+4))\bigr]\, g = 0   \,,
\label{Ecg(t)}
\end{equation}
that is, a new Gauss hypergeometric equation
$$
 t\,(1 - t)\,g_{tt}'' + [c-(1+a_\k'+b_\k')t]\,g_t' - a_\k' b_\k' g = 0 \,,
$$
so that  the solution regular at $t=0$ is the hypergeometric function
$$
 g(t,\k) = {}_2 F_1(a_\k',b_\k';c\,; t) \,,\qquad
 {}_2 F_1(a_\k',b_\k';c\,; t) = 1 + \sum_{n=1}^{\infty}\,
 \frac{(a_\k')_n\,(b_\k')_n}{(c)_n} \,\frac{t^n}{n\,!} \,,
$$
with $a_\k'$ and $b_\k'$ given by
$$
 a_\k' = \frac{1}{2}\Bigl[(L+2) \pm B_\k  \Bigr]  \,,{\quad}
 b_\k' = \frac{1}{2}\Bigl[(L+2) \mp B_\k  \Bigr] \,,{\quad}
 B_\k = \frac{\sqrt{\wt{\k}(\wt{\k}+1)}}{\wt{\k}}  \,.
$$
The polynomial solutions  appear when one of the two $\k$-dependent coefficients, $a_\k'$ or $b_\k'$, coincides with zero or with a negative integer number
$$
 a_\k' = -\,n_r\,,\quad {\rm or}\quad
 b_\k' = -\,n_r\,,\qquad  n_r=0,1,2,\dots
$$
In this case, we arrive to the following expression for $ \wt{\k}$:
$$
  \wt{\k} =\k/{\cal E}^2  =  1/((2 n_r + L + 1) ( 2 n_r + L + 3))  \,,
$$
that leads to the the following expression for discrete values of the energy:
$$
 {\cal E}_{{n_r}l}^2 = \k\, [(2 n_r + L + 1) ( 2 n_r + L + 3)]
=   \k\, [(2 n_r+L+2)^2-1]  \,. 
$$

The recurrence relation leads to the following recursions for the even coefficients
$$
 g_2 = \frac{\k (L+1)(L+3) - {\cal E}^2}{2 (2 L + 3)}\,\,g_0 \,,\quad
 g_4 = \frac{\k (L+3)(L+5) - {\cal E}^2}{4 (2 L + 5)}\,\,g_2  \,,\quad
 g_6 = \frac{\k (L+5)(L+7) - {\cal E}^2}{6 (2 L + 7)}\,\,g_4 \,,\quad \dots
$$
so that the function $g$ is given by
\begin{eqnarray*}
 g &=&  1 + \frac{\k (L+1)(L+3) - {\cal E}^2}{2 (2 L + 3)}\,s^2
  + \frac{[\k (L+1)(L+3) - {\cal E}^2][\k (L+3)(L+5) - {\cal E}^2]}{8 (2 L + 3)(2 L + 5)}\,s^4 \cr
 &+&      \frac{[\k (L+1)(L+3) - {\cal E}^2][\k (L+3)(L+5) - {\cal E}^2][\k (L+5)(L+7) - {\cal E}^2]}{48 (2 L + 3)(2 L + 5)(2 L + 7)}\,s^6  + \dots
 \end{eqnarray*}
\begin{enumerate}
\item  Suppose that $n_r=0$; then ${\cal E}^2 = {\cal E}_{0L}^2 = k (L+1) (L+3)$
and $g_2=0$. In this particular case, the function $g$ reduces to
$$
  {\cal P}_0^{\,g} =  1 \,.
$$
\item  Suppose that $n_r=1$; then ${\cal E}^2 = {\cal E}_{1L}^2 = k
(L+3)(L+5)$ and $g_4=0$. In this particular case, the function $g$
reduces to the polynomial
$$
  {\cal P}_{1L}^{\,g} =   1 - \frac{2 (L+3)}{(2 L + 3)}\, \k\s^2 \,.
$$

\item  Suppose that $n_r=2$; then ${\cal E}^2 = {\cal E}_{2L}^2 = k
(L+5)(L+7)$ and $g_6=0$.  In this particular case, the function $g$
reduces to the polynomial
$$
  {\cal P}_{2L}^{\,g} = 1 - \frac{4 (L+4)}{(2 L + 3)}\, \k\s^2 + \frac{4 (L+5)(L+4)}{(2 L + 3)(2 L + 5)}\, \k^2\s^4\,.
$$

\item In the general case for any $n_r$, then $\caE^2 = \caE_{n_rL}^2 = \k\, (2 n_r + L + 1) (2 n_r + L + 3)$, $n_r=0,1,2,\dots$, and  then $g_{2n_r} \ne 0$, but $g_{2n_r+2}=0$. The function $g$ reduces to  a polynomial ${\caP}_{n_rL}^{\,g}$ of degree $2n_r$ in the variable $\s$, with only even powers.

\end{enumerate}

The following proposition is similar to Proposition 2. 
%--------------------
%%  {Proposicion 3}
\begin{proposicion}
The eigenfunctions corresponding to distinct eigenvalues of the
Sturm-Liouville problem given by the equation
$$
 a_0  g'' + a_1 g' + a_2 g = 0  \,,
$$
with
$$
 a_0 = \s\,(1 - k\,\s^2) \,,{\quad}  a_1 =  2 (L+1) -  \k (2L + 5) \s^2    \,,{\quad}
 a_2 = \bigl[ \caE^2 - 3\k -  \k L (L+4)\bigr]\,\s\,,
$$
together with the appropriate boundary conditions in the points
$\s_1=0$ and $\s_2=1/\sqrt{\k}$  are orthogonal in the interval
$[0,1/\sqrt{\k}]$ with respect to the weight function $q_g =
\s^{2(L+1)} \sqrt{1 - \k\,\s^2}$.
\end{proposicion}

Notice that this problem is also singular (as in Propsition 2) and that the weight function $q_g$  is different from the one in the type I solutions. The eigenfunctions are just the polynomial solutions ${\caP}_{n_rL}^{\,g}(s)$, $n_r=0,1,2,\dots$ previously obtained that satisfy the orthogonality relations
$$
  \int_{0}^{1/\sqrt{\k}} {\caP}_{n_{r_1}L}^{\,g}(\s,\k)\,{\caP}_{n_{r_2}L}^{\,g}(\s,\k)\, q_g(r,\k)\, d\s = 0
  \,,\quad n_{r_1}\,\ne\,n_{r_2} \,,\quad \k>0\,,
$$
that can be rewritten as follows
$$
  \int_{0}^{1/\sqrt{\k}} R_{n_{r_1}L}^{\,g}(\s,\k)\,R_{n_{r_2}L}^{\,g}(\s,\k)\,\,
  \Bigl(\frac{\s^2}{\sqrt{1-\k\,\s^2}}\Bigr)\,d\s  = 0   \,,\quad
  n_{r_1}\,\ne\,n_{r_2} \,, 
$$
where $R_{n_{r_1}L}^{\,g}$ and $R_{n_{r_2}L}^{\,g}$  are the radial functions 
$$
  R_{n_{r_1}L}^{\,g} = \s^L\,\sqrt{1 - \k\,\s^2}\,\,{\caP}_{n_{r_1}L}^{\,g}(\s,\k)  
  \,,{\qquad} 
  R_{n_{r_2}L}^{\,g} = \s^L\,\sqrt{1 - \k\,\s^2}\,\,{\caP}_{n_{r_2}L}^{\,g}(\s,\k) \,. 
 $$

%--------------------------------------
%%  (sub-Section 4.3)
\subsection{Final solution for the sphere and its Euclidean limit}

By working in the coordinate chart $(s, \te, \phi)$, which covers the upper hemisphere, we have obtained the following  solutions for the radial function $R(\s)$ in (\ref{sFactorization}) (in each case, the quantum numbers $n_r$ and $L$ are independent and their ranges are $n_r=0, 1, 2, \dots$ and $L=0, 1, 2\dots$; remark also that had we worked in the alternative $(s, \te, \phi)$ chart covering the lower hemisphere, the expressions for the solutions would have been the same): 
\begin{itemize}
\item[(I)]  Solutions of type I : The radial functions are of the form  
$$  
  R = \s^L {\caP}_{n_r L}^{\,f}(\s,\k)  \,,
$$
and the value of the associated energy, with quantum numbers $n_r$ and $L$,  is given by
$$
 {\cal E}_{{n_r}L}^2 = \k\, (2 n_r + L) ( 2 n_r + L + 2) = \k\, N\, (N + 2)\,,\ 
 N  = 2 n_r+L. 
$$
\item[(II)]  Solutions of type II : The radial functions are of the form 
$$  
  R =  \s^L\,\sqrt{1 - \k\,\s^2}\,\,{\caP}_{n_rL}^{\,g}(\s,\k) \,, 
$$
and the value of the associated energy, with quantum numbers $n_r$ and $L$,  is given by 
$$
 {\cal E}_{{n_r}L}^2 = \k\, (2 n_r + L + 1) ( 2 n_r + L + 3) = \k\,N (N + 2)\,,\  N  = 2 n_r+1+L. 
$$
\end{itemize}

Two important properties are (i) in both cases ${\cal E}^2$ depends only on a single  total quantum number $N$, so the energy levels are degenerate with respect $n_r$ and $L$ (each level comprises solutions of type I and type II), and  (ii) in both cases, ${\cal E}^2$ is proportional to the curvature.

We have found that in the $\k>0$ case, the free particle  (which can be considered as a very  special case of a central potential) is also endowed with extra accidental degeneracy, further to the rotational degeneracy. We also note that the expression of $N$ as a function of $n_r$ and $L$ is the same that appears in the analogous problem for the harmonic oscillator. 

Now, we can put together all the previous results  and rewrite them in terms of the usual geodesic coordinates $(r, \te, \phi)$. Apart of the known singularities in the angular part, these geodesic coordinates cover the whole sphere, with radial singularities only at $r=0$ (the origin, or North pole) and at $r=\pi/(2\sqrt{\k})$ (the antipodal point, or South pole). Now, as in both charts $\s=\Sin_\k(\r)$, it turns out that $\s^2 = (1- \Cos_\k(\r))/\k$, so that in the upper hemisphere $\Cos_\k(\r) = \sqrt{1-\k s^2}$ and on the lower one $\Cos_\k(\r) = -\sqrt{1-\k s^2}$. By reexpressing either 
the polynomials ${\caP}_{n_r L}^{\,f}(\s,\k)$ (I below)  or the function $\sqrt{1 - \k\,\s^2}\,\,{\caP}_{n_rL}^{\,g}(\s,\k)$ (II below) in terms of the variable $\xi\equiv\Cos_\k(\r))$, we get a family of polynomials denoted by $\caS_{nL}(\xi) \equiv \caS_{nL}(\Cos_\k(\r))$ whose degrees are
\begin{itemize}
\item[I]  $n=2n_r$, so $n$ is even for the polynomials $\caS_{nL}(\xi)$ coming from type I solutions.  
\item[II] $n=2n_r+1$, so $n$ is odd for the polynomials $\caS_{nL}(\xi)$ coming from type II solutions.   
\end{itemize}
and with coefficients wich are $\k$-independent. 
It is clear from the previous analysis that the polynomials $\caS_{nL}$ of degree $n$ in the variable $\xi\equiv\Cos_\k(\r))$ provide a solution of our problem for all values of the radial coordinate $r$, this is, in both hemispheres. The symmetry of the sphere around its equator, conveyed by the transformation $r \to \pi/\sqrt{\k} -r$, reflects itself in the even character of the polynomials $\caS_{nL}$ with $n$ even and the antisymmetry of those with $n$ odd under the replacement $\xi \to -\xi$; notice that both type I and type II solutions are encompassed under a single family of polynomials $\caS _{nL}(\xi)$.

The  polynomials $\caS _{nL}(\xi)$, for the first low values for $n=0, 1, 2, 3, 4, 5, 6$ and arbitrary $L$ are given by 
\begin{eqnarray*}
 \caS_{0L}(\xi) &=& 1\\[2pt]
 \caS_{1L}(\xi) &=& \xi\\[2pt]
 \caS_{2L}(\xi)  &=& k_2\,\left( -1 + 2(L+2) \xi^2\right)\\[2pt]
 \caS_{3L}(\xi)  &=& k_3\,\left( -3\xi + 2(L+3) \xi^3\right)\\[2pt]
 \caS_{4L}(\xi)  &=& k_4\,
     \left( 3 -12(L+3) \xi^2 + 4(L+3)(L+4)\xi^4\right)\\[2pt]
 \caS_{5L}(\xi) &=& k_5\,
     \left( 15\xi -20(L+4) \xi^3 + 4(L+4)(L+5)\xi^5\right)\\[2pt]
 \caS_{6L}(\xi)  &=& k_6\, 
    \left( -15 + 90 (L+4) \xi^2 -60(L+4)(L+5)\xi^4 + 8 (L+4)(L+5)(L+6)\xi^6\right)
\end{eqnarray*}
with the following values for the global coefficients  
$$
 k_2 = k_3 = \frac{1}{2L + 3}  \,,\quad 
 k_4 = k_5 = \frac{1}{(2L + 3)(2L+5)}  \,,\quad  
 k_6 =   \frac{1}{(2L + 3)(2L+5)(2L+7)}  \,, 
$$
and in terms of these polynomials, the  wavefunction corresponding to a spherical wave on the sphere, depending of the quantum numbers
$(n,L,m)$ is given in geodesic spherical coordinates $(r,\te,\phi)$ by
$$
 \Psi_{n,L,m} = (\Sin_\k(\r))^L {\caS}_{n,L}(\Cos_\k(\r))\, \,Y_{Lm}(\te,\phi)
  \,,{\quad}  n=0,1,2,\dots
$$
Notice the analogy for the radial dependence of these solutions with the angular dependence in the angular coordinate $\te$ involving the Legendre polynomials with the variable $\cos\te$ and the associated Legendre functions involving extra powers of $\sin\te$. This can be expected as, after all, the problem we are discussing is finding  eigenfunctions of the Laplacian in the sphere $S^3$, which leads to the hyperspherical harmonics; in fact, the polynomials ${\caS}_{n,L}(\xi)$ are proportional to the Gegenbauer polynomial $C_{n}^{L+1}(\xi)$. 

We state together several properties of the polynomials ${\caS}_{n,L}(\xi)$, some of which have been already mentioned, while the rest can be easily derived:

\begin{enumerate}
\item ${\caS}_{n,L}(\xi)$ is a polynomial of degree $n$ (which is even/odd for even/odd values of $n$). All its roots are real and are contained in the interval $[-1,1]$ and as they come in pairs $\lambda, -\lambda$, there are precisely $[n/2]$ roots in the interval $[0,1]$ and an extra root at $\xi=0$ when $n$ is odd (Here, $[n/2]$ denotes the integer part of $n/2$). 

\item At the North pole, where $\xi=1$ and for all $n, L$, we have ${\caS}_{n,L}(1) = 1$; as the polynomial is even/odd, this implies that at the South pole, where $\xi=-1$ we have ${\caS}_{n_r,L}(-1) = \pm 1$ according as $n$ is even / odd, and this for all $L$. 

\item At the equator, $\xi=0$, for $n$ even, the value of the polynomial ${\caS}_{n,L}(0)$ is always different from zero (this value is positive for $[n/2]$ even  and negative for $[n/2]$ odd, and the absolute value of ${\caS}_{n,L}(0)$ decreases when $n$ or $L$ grows). For $n$ odd the value of the polynomial ${\caS}_{n,L}(0)$ is always equal to zero

\end{enumerate}

The orthogonality relations previously found, when written in terms of the radial variable $r$ and of the polynomials ${\caS_{n,L}}(\Cos_\k(\r))$, (with $n$ either even or odd) become 
$$
  \int_{0}^{\pi/\sqrt{\k}} \Bigl[\Sin_\k(\r)^L {\caS_{{n_1},L}}(\Cos_\k(\r))\Bigr]\,   \Bigl[\Sin_\k(\r)^L{\caS_{{n_2},L}}(\Cos_\k(\r))\Bigr]
  \Sin_\k(\r)^2\,dr  = 0   \,,{\quad}
  n_{1}\,\ne\,n_{2} \,,{\quad}  \k>0\,.
$$
which is a bit more general than the previous expressions, as it includes solutions of type I and II altogether.

A few polynomials with $n=4,5,12,13$ and several values of $L$ are displayed in the figures 1 to 4.

  To sum up, the wave functions  of the free particle in the
three-dimensional sphere  $S_{\k}^3$, which have a separated expression in 
the geodesic spherical coordinates $(r,\te,\phi)$, depend on the quantum numbers
$(n,L,m)$ and are given (up to a multiplicative constant) by
$$
 \Psi_{n,L,m} \propto  (\Sin_\k(\r))^L {\caS}_{n,L}(\Cos_\k(\r))\, \,Y_{Lm}(\te,\phi)
  \,,{\quad}  n=0, 1,2,\dots,
$$
with energies given by
$$
  E_{n,L}= \left(\frac{\hbar^2}{2m}\right)\k\, (n + L) (n + L+ 2)  \,,{\quad} n=0, 1,2,\dots, \quad L = 0, 1, 2, \dots
$$
that are proportional to the curvature $\k$ and depend only of a
total quantum number $N$ given by $N = n + L = 0, 1, 2, \dots$. Further to the rotational degeneracy, 
contained in a degeneracy factor $2L+1$, the levels in this problem have an extra 
degeneracy wich leads to a total degeneracy $(N+1)^2$ for the level with total quantum number $N$ (e.g., for $N=4$ we have $4= 4 + 0 = 3 + 1 = 2 + 2 = 1 + 3 = 0 + 4$, with total degeneracy $1 + 3 + 5 + 7 + 9  = 25$, etc.) 

Another question to be discussed here refers to the $\k\to 0$ limit of 
the sphere results, where the known Euclidean results should be recovered. 
We recall briefly these Euclidean results: 
The wave functions of the free particle in the three-dimensional 
Euclidean  $\IE^3$, which have a separated expression in 
the geodesic spherical coordinates $(r,\te,\phi)$ are the {\itshape spherical waves} that depend on a continuous positive label $k$ and two quantum numbers
$(L,m)$, and are given (up to a multiplicative constant) together with their energies by
$$
 \Psi_{k;L,m} \propto  j_L(k r)\, \,Y_{Lm}(\te,\phi),   {\qquad}
  E_{k}= \left(\frac{\hbar^2}{2m}\right) k^2,
$$
where $j_L(k r)$ are the spherical Bessel functions. 

The limit from the sphere $S_{\k}^3$ to the Euclidean space $\IE^3$, requires some care. In the polynomials ${\caS}_{n,L}(\Cos_\k(\r))$, whose coefficients are independent of $\k$, all the dependence on the curvature is contained in the argument $\Cos_\k(\r)$ which in the limit $\k\to 0$ goes to $1$ for any value of $r$, so in the simple $\k\to 0$ limit the polynomial reduces to a constant. In the same `naive' limit, with $\k\to 0$ and fixed quantum numbers, the energy would go to cero. Physically, what one should obtain is a continuous spectrum for the energy, bounded from below by the value $E=0$; this  cannot be done simply by means of $\k\to 0$, but would require to maintain the value of $\caE$ in (\ref{E(nrL)}), what implies that $n\to\infty$ should be made at the same time as $\k\to0$ (the quantum number $L$ is kept fixed in this proccess).  In other words, the simple limit $\k\to0$ would make the quantum label $n$ to disappear, without leaving any trace, but in the true limit, there is another quantum label $k$, which should appear.  This mean that the limit should be made enforcing the constancy of $\k\, [(n+L+1)^2-1]$ throughout. 

When this is duly  taken into account, everything fits. Of course, this is a consequence of the relation between  the radial equation for the sphere (\ref{EcR(rho1)}) which is a kind of deformed spherical Bessel equation  and its Euclidean limit (\ref{EcR(rho2)}) which is precisely the spherical Bessel equation. It is interesting to look at this limit, where the complete Euclidean space appears as the limit of the {\itshape upper half-sphere}, with the sphere equator going to the Euclidean infinity (this fits with the use of the chart $(s, \te, \phi)$ which in the limit covers the complete Euclidean space). Consequently, the complete radial behaviour of the Euclidean radial function comes from the `upper half', from $\xi=1$ to $\xi=0$ of the full interval $[-1, 1]$ for $(\Cos_\k(\r))$, with the equator (with $r=\pi/(2\sqrt{\k})$ going to the Euclidean infinity as $\k\to0$. This corresponds to the limit
\begin{equation}
\lim_{\k\to 0;\  n\to\infty} (\Sin_\k(\r))^L {\caS}_{n,L}(\Cos_\k(\r)) \propto  j_L(k r)
\label{S3ToE3Limit}
\end{equation}
whenever the control variables $\k$ (continuous and positive) and $n$ (discrete) are linked throughout the limit by the condition $\k\, (n + L) (n + L+ 2) = k^2$. The precise proportionality factor required to turn (\ref{S3ToE3Limit}) into an equality comes from matching the lowest coefficients: for small $r$, $(\Sin_\k(\r))^L {\caS}_{n_r,L}(\Cos_\k(\r))\approx r^L$, while for small $x$, $j_L(x)\approx x^L/(1 \cdot 3\cdot 5\cdot \cdot (2L+1))$. 

In Figures 5 and 6 we display this fact, approaching the spherical Bessel functions quite closely with only a few members of the limit sequence indicated above.  The concordance is good enough with $n$ moderately low, which allows us to avoid the computational instabilities ---catastrophic cancellation due to the large absolute values and alternating signs of the coefficient for the polynomials--- which arises for higher values of $n$.

%--------------------------------------
%% (Section 5.)
\section{Hyperbolic $\k<0$ case}

The search of the analogues of the spherical waves in the hyperbolic space can be carried out by using the same approach.   However,  there are some relevant differences. 
Thus we restrict here to a sketch of the main traits, which will be discussed in full detail elsewhere. 

In this case, we have  $\k=-|\k|<0$ and by using the adimensional variables defined similarly to the sphere case, the equations
(\ref{EcR(rho1)}) and (\ref{Ecf(rho1)}) become
\begin{equation}
 {\rho^2} (1 + |\wt{\k}|\,\rho^2)\,R''
 + \rho(2 + 3 |\wt{\k}|\,\rho^2)\,R' +\bigl( \rho^2 - L(L+1) \bigr)\,R= 0 \,,
\label{EcR(rhoh)}
\end{equation}
and
\begin{equation}
 \rho\,(1 + |\wt{\k}|\,\rho^2)\,f'' + \bigl[2 (L+1) + |\wt{\k}| (2L + 3)\rho^2 ]\,f'
 + \bigl[ 1 + |\wt{\k}| L (L+2)\bigr]\,\rho\,f = 0  \,,
\label{Ecf(rhoh)}
\end{equation}
%%  \begin{equation}
%%    4 z\,(1 + |\wt{\k}|\,z)\, f_{zz}'' + 2 [ (2L+3)  + 2 |\wt{\k}| (L+2)z]\,f_z' + [1 + |\wt{\k}| L (L+2)] f = 0  \,.
%%  \label{Ecf(zk1h)}
%%  \end{equation}
that also represent a deformation of the Bessel equation and
that can also be solved by using a power series procedure. The
main difference is that now the singularity of the equation is 
placed on the $\rho$ imaginary axis and the equation is
well defined for all the real values $\rho>0$.  This fits of course 
to the fact that in the hyperbolic space, the range of the geodesic 
radial distance $r$ is $[0, \infty)$ and this implies also that  
both $s$ and  $\rho$ have the range $[0, \infty)$. The substitution
$z=\rho^2$ and then  the change  $t=-|\wt{\k}|\, z$ leads to the
following hypergeometric equation:
$$
 t\,(1 - t)\,f_{tt}'' + \Bigl[(L+\frac{3}{2}) - (L+2)\,t\Bigr]\,f_t'
 - \frac{1}{4|\wt{\k}|}\bigl[1 + |\wt{\k}|\,L\,(L+2)\, \bigr]\,f  = 0   \,,
$$
which nevertheless has a singularity at $t=1$, which in this case comes from the singularity at the imaginary axis through the changes from $\rho$ to $z$ and to $t$. 

This equation  is a Gauss hypergeometric equation with $c=L+3/2$ and
$a_\k$ and $b_\k$ given by
$$
 a_\k = \frac{1}{2}\Bigl[(L+1) \pm B_\k  \Bigr]  \,,{\quad}
 b_\k = \frac{1}{2}\Bigl[(L+1) \mp B_\k  \Bigr] \,,{\quad}
 B_\k = \frac{\sqrt{|\wt{\k}|^2 - |\wt{\k}|}}{|\wt{\k}|}  \,.
$$
So the solution of the quantum free particle in a hyperbolic
space can also be expressed in terms of a hypergeometric series that
can be considered as representing a $\k$-deformation of the
Euclidean spherical waves.

We close this section with two points that must be remarked.
First, in the case of the hyperbolic space this hypergeometric 
equation does not admit polynomial solutions since  the condition 
for $a_\k$ or $b_\k$ to coincide with zero or a negative integer 
number leads to $|\k|<0$ (a condition impossible to be satisfied). 
This fits with the physical expectations that the spectrum
for the energy $E= (\hbar^2/2m)\caE^2$ would be continuous. 
Second, for certain values of the curvature the coefficients $a_\k$ and $b_\k$
become complex; nevertheless, the point is that they are complex conjugated (that is, $b_\k^*=a_\k$) and because of this the solution (that is a hypergeometric function) 
continues to be a real function. In fact, we have obtained two properties (no polynomial solutions and existence of complex coefficients) that
coincide with similar results obtained (but with a different
formalism) in the previous study \cite{CaRaSa11a} of the
two-dimensional system.

%--------------------------------------
%% (Section 6.)
\section{Final comments and outlook}

We have studied the analogous to the Euclidean spherical waves for a 
quantum free particle on the three-dimensional spherical and hyperbolic spaces using a curvature dependent approach. 
We start from  the $\k$-dependent Killing vectors of the metric and we obtain the quantum (kinetic) Hamiltonian $\widehat{H}(\k)$ as a function of the operators $\widehat{P_i}$, and  $\widehat{J_i}$. 
Then the $\k$-dependent Schr\"odinger equation (that when $\k\to 0$ reduces to a spherical Bessel differential equation) can be carried in the case $\k\neq 0$  through a number of changes of variable, to a hypergeometric equation. 
We study with full detail the case for positive curvature on a sphere, where ---as the space is closed--- quantum mechanics leads naturally to a discrete spectrum for the energy of the analogous of spherical waves.  
In the standard Euclidean quantum mechanics, this discretization is usually a consequence of the existence of a potential. In the spherical $\k>0$ case, even without
the presence of a potential, the geometry of the space  produces a
discrete spectrum. We also discuss in detail the limit from the sphere to the Euclidean space. 

We call the attention to some particular points.  
First, the differences between the spherical $\k>0$ and the hyperbolic
$\k<0$ cases have been clearly stated; nevertheless, this is a point deserving to be studied with more detail.   Second, in the spherical $\k>0$ case, one can expect some specific relations to exists between $\k$-dependent plane-waves (obtained in \cite{CaRaSa11a}) and the spherical waves obtained here (in the Euclidean case this a well known relation), and to find this explicitly can be considered as an open question.  
Third, in the $\k<0$ hyperbolic case (that, in certain aspects, seems more similar to the Euclidean case that the spherical one) the equations can lead to complex values for the parameters $a_\k$ and $b_\k$; this  also deserves be studied. 

We finalize with two interesting open questions: 

\begin{itemize}

\item[(i)] Finally, in the $\k>0$ case we have obtained a family of orthogonal polynomials (that is different to the family obtained in \cite{CaRaSa11a}). We have seen in some previous papers that this $\k$-formalism provides, as a byproduct, new families of polynomials; as an example, in \cite{CRS07AnPh1} were obtained the so-called curved Hermite polynomials (CHP). We think that the family here obtained deserves also to be studied (that is, existence of the corresponding `deformed' Rodrigues formula, generating function,  or recursion relations) probably in a rather similar way as the one presented in  \cite{ViLa09}.

\item[(ii)]  The quantum  free motion on the 3-dim spherical and hyperbolic spaces  should correspond to the limiting case of the Hydrogem atom (see e.g. \cite{Sch40}) or the harmonic oscillator (see e.g. \cite{CRS12}) in curved spaces when the coupling constant in the potential vanishes. We note that in these two cases the problem involves two parameters (coupling constant and curvature) and there are therefore two different limiting procedures. A general analysis of these limits is a delicate problem that deserves to be studied.

\end{itemize}

%--------------------------------------
\section*{Acknowledgments}

The  authors are indebted to the referee for some interesting remarks  which have improved the presentation of this paper.   
 JFC and MFR acknowledges support from research projects MTM--2009--11154
(MCI, Madrid)  and DGA-E24/1 (DGA, Spain); and MS from research projects MTM--2009-10751 (MCI, Madrid).
%%%   and JCyL-GR224-08 (JCyL, Spain).

{\small

%--------------------------------------
    }
%--------------------------------------
%---------------
\vfill\eject
%---------------
%--------------------
%% Figure 1
\begin{figure}\centerline{
\scalebox{1.3}{\includegraphics{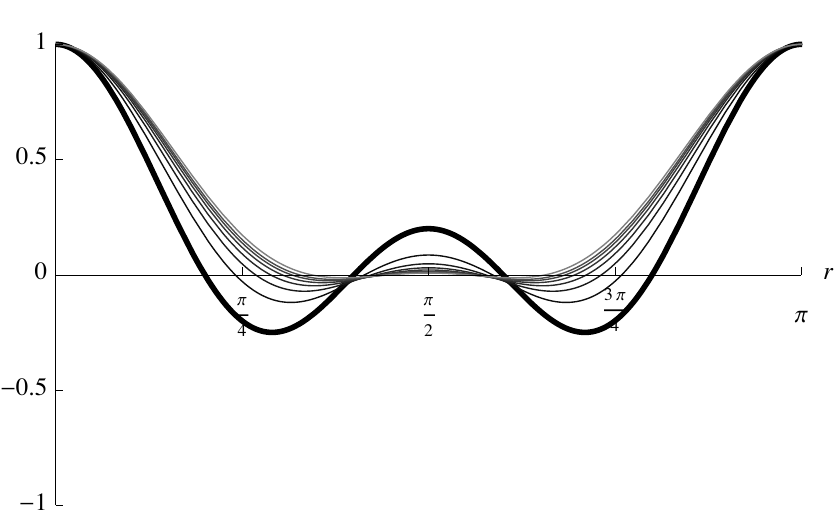}} }

\caption{Several polynomials ${\caS_{n,L}}(\Cos_\k(\r))$ with $n=4$ and values of $L=0, 1, 2, 3, 4, 5, 6, 8$. The polynomial with $L=0$ is in dark grey, thick, and increasing $L$ corresponds to lighter grey. The variable $r$ in abscissas is the geodesic lenght on a sphere with curvature $\k=1$, so $r=0$ corresponds to the North Pole and $r=\pi$ corresponds to the South pole. Notice all these polynomials have precisely 4 zeros and all are even around the equator, $r=\pi/2$.}
\label{Fig1}
\end{figure}

%--------------------
%% Figure 2
\begin{figure}\centerline{
\scalebox{1.3}{\includegraphics{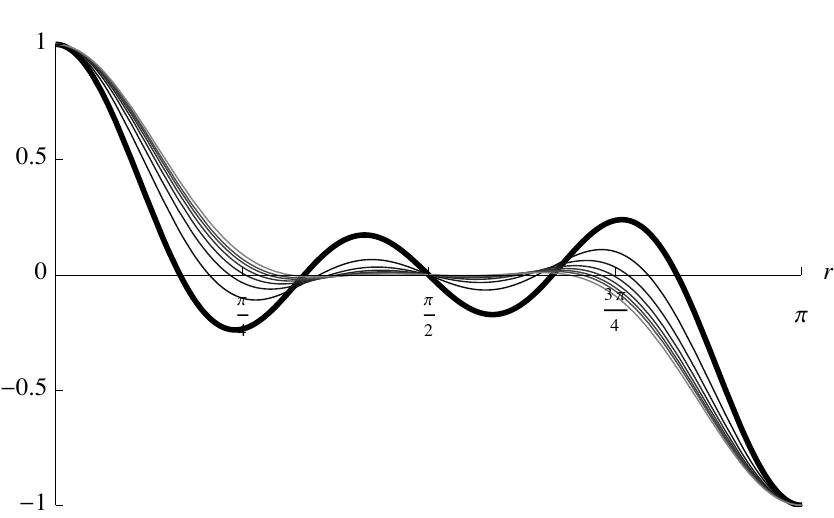}} }

\caption{Several polynomials ${\caS_{n,L}}(\Cos_\k(\r))$ with $n=5$ and values of $L=0, 1, 2, 3, 4, 5, 6, 8$. The polynomial with $L=0$ is in dark grey, thick. The variable $r$ in abscissas is the geodesic lenght on a sphere with curvature $\k=1$, so $r=0$ corresponds to the North Pole and $r=\pi$ corresponds to the South pole. Notice all these polynomials have precisely 5 zeros and all are odd around the equator, $r=\pi/2$}
\label{Fig2}
\end{figure}

%---------------
\vfill\eject
%---------------

%--------------------
%% Figure 3
\begin{figure}\centerline{
\scalebox{1.3}{\includegraphics{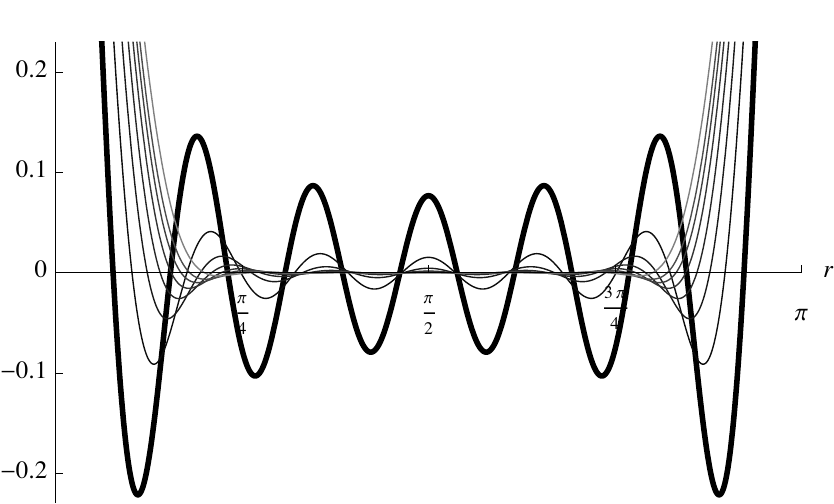}} }

\caption{Several polynomials ${\caS_{n,L}}(\Cos_\k(\r))$ with $n=12$ and values of $L=0, 1, 2, 3, 4, 5, 6, 8$. Same conventions as in preceeding figures. Notice all these polynomials have precisely 12 zeros and all are even around the equator, $r=\pi/2$; the amplitude of oscillations at the equatorial band decreases when $L$ grows, and to allow this to be seen clearly the range displayed has been reduced.}
\label{Fig3}
\end{figure}

%--------------------
%% Figure 4
\begin{figure}\centerline{
\scalebox{1.3}{\includegraphics{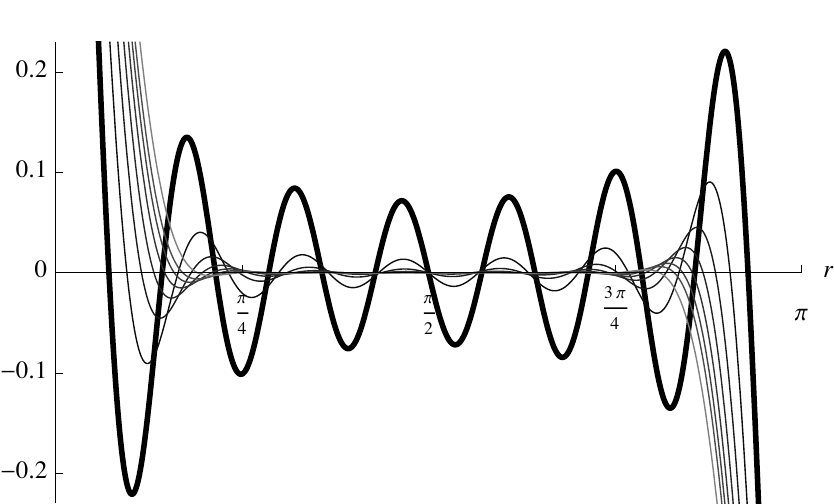}} }

\caption{Several polynomials ${\caS_{n,L}}(\Cos_\k(\r))$ with $n=13$ and values of $L=0, 1, 2, 3, 4, 5, 6, 8$. Notice all these polynomials have precisely 13 zeros and all are odd around the equator, $r=\pi/2$.}
\label{Fig4}
\end{figure}

%---------------
\vfill\eject
%---------------

%--------------------
%% Figure 5
\begin{figure}\centerline{
\scalebox{1.3}{\includegraphics{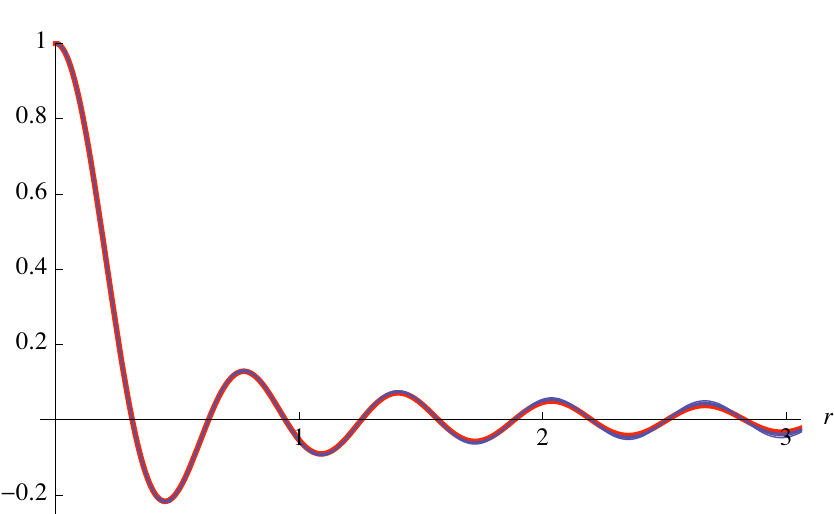}} }

\caption{Aproaching the Euclidean radial function for $L=0$ (the spherical Bessel function $j_0(kr)$ for $k=10$ with a sequence of the solutions for the sphere, corresponding to the values $n=20, 24, 32, 40$ and $L=0$. In the limit $\k\to 0$ the upper hemisphere goes to the full Euclidean plane and the sphere Equator goes to Euclidean infinity, so here only the left hand of the graphics in figures 1 to 4 (which corresponds to the upper hemisphere) is pertinent.}
\label{Fig5}
\end{figure}

%--------------------
%% Figure 6
\begin{figure}\centerline{
\scalebox{1.3}{\includegraphics{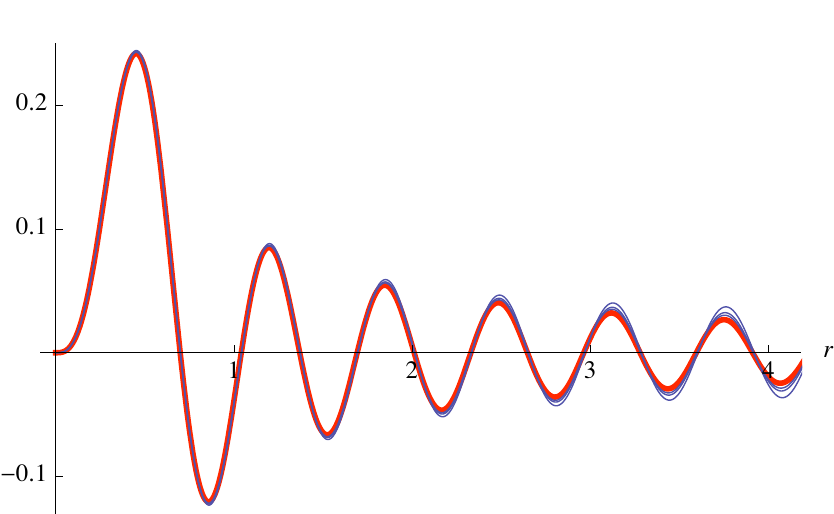}} }

\caption{Same as Figure 5 but for $L=3$, approaching the spherical Bessel function $j_3(kr)$.}
\label{Fig6}
\end{figure}

%---------------
\end{document}